\documentclass[journal=ancac3,manuscript=article]{achemso}

\usepackage{chemformula} 
\usepackage[T1]{fontenc} 
\usepackage{soul, color}
\usepackage{wasysym}
\usepackage{subcaption}
\usepackage{natbib}
\usepackage{mathtools}

\author{Robert L. Stamps}
\email{Robert.Stamps@umanitoba.ca}
\affiliation{Department of Physics \& Astronomy, University of Manitoba, Winnipeg, Canada}

\author{Rehana Begum Popy}
\affiliation{Department of Physics \& Astronomy, University of Manitoba, Winnipeg, Canada}

\author{Johan van Lierop}
\email{Johan.van.Lierop@umanitoba.ca}
\affiliation{Department of Physics \& Astronomy, University of Manitoba, Winnipeg, Canada}

\title{Active Inference and Artificial Spin Ice: Control Processes and State Selection}


\begin{document}

\begin{tocentry}

%
%
%

\centering
\includegraphics[scale=0.45]{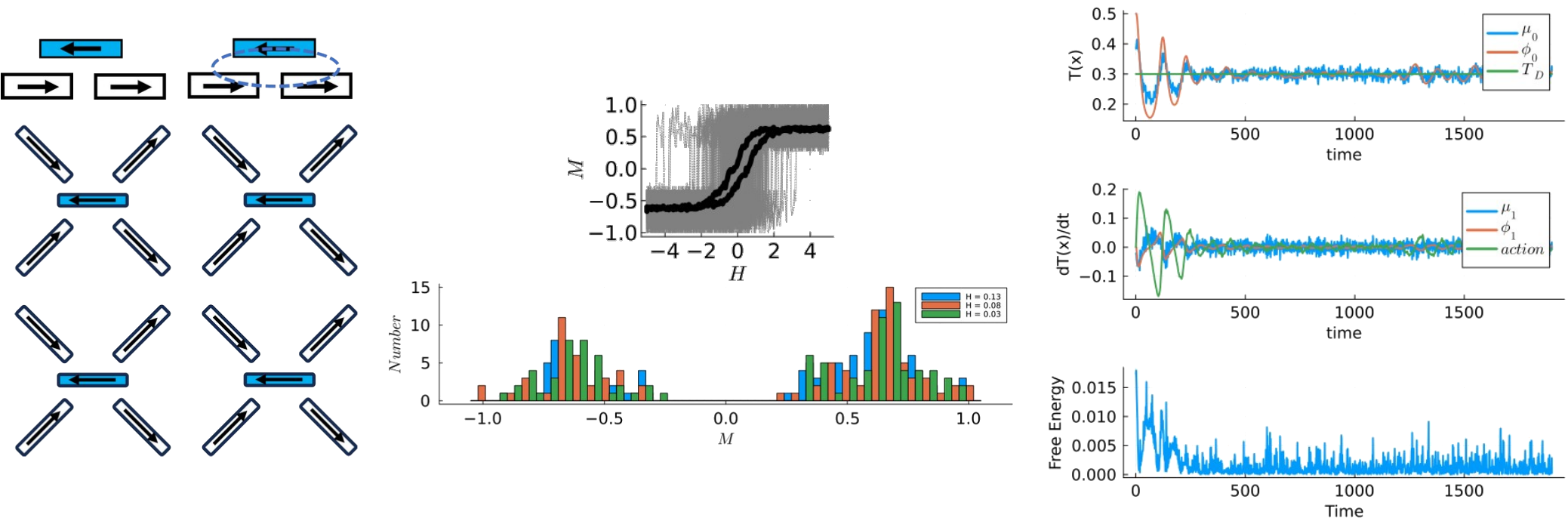}

\end{tocentry}

\begin{abstract}
A numerical model of interacting nanomagnetic elements is used to demonstrate active inference with a three dimensional Artificial Spin Ice structure.  It is shown that thermal fluctuations can drive this magnetic spin system to evolve under dynamic constraints imposed through interactions with an external environment as predicted by the neurological free energy principle and active inference. The structure is defined by two layers of magnetic nanoelements where one layer is a square Artificial Spin Ice geometry. The other magnetic layer functions as a sensory filter that mediates interaction between the external environment and the ``hidden'' Artificial Spin Ice layer. Spin dynamics displayed by the bilayer structure are shown to be well described using a continuous form of a neurological free energy principle that has been previously proposed as a high level description of certain biological neural processes. Numerical simulations demonstrate that this proposed bilayer geometry is able to reproduce theoretical results derived previously for examples of active inference in neurological contexts.
\end{abstract}

In recent years new insights into how a brain might processes and respond to information have emerged from a neurological theory whose primary components are called active inference and the free energy principle. The theory provides a high-level mathematical description of neural processes where comparisons are made between predictions and sensory perceptions resulting in adjustments that affect future sensory input. The theory is built from simple general assumptions and has been proposed as a biologically plausible mechanism that can describe key aspects of motor control and movement regulation.~\cite{friston_free_2006,adams_active_2015,friston_dem_2008, friston_generalised_2010,friston_free-energy_2010} A fundamental hypothesis is that many brain functions can be described by Bayesian inference that can be expressed mathematically using a variational Bayes technique.~\cite{friston_free-energy_2009,aguilera_how_2022,fitzgerald_active_2015} 

The purpose of the present paper is to show that the same theory can describe dynamics of a non-biological system and how this system might be studied experimentally. Moreover, we show that active inference in a physical device that permits detailed experimental study would open up new avenues of investigation that provide a new methodology for probing complex out-of-equilibrium dynamics and would also have significant potential for a new type of physical neuromorphic computing\cite{Nakajima.2020}.  

We demonstrate these ideas with a model based on Artificial Spin Ice (ASI). Research into ASI was initially directed towards investigations of complex frustrated systems with an emphasis on experiment.~\cite{skjaervo_advances_2020,nisoli_colloquium_2013} One of the  benefits has been to provide experimental systems that can be probed with unprecedented detail on time- and length-scales that are not otherwise amenable to study.~\cite{marrows_experimental_2021} Some of the most recent developments have expanded the field to include neuromorphic applications for machine learning\cite{jensen_reservoir_2020,hon_numerical_2021, gartside_reconfigurable_2022} and fabrication of non-trivial three dimensional structures in complex geometries~\cite{may_magnetic_2021,saccone_exploring_2023}. 

\section{Nanomagnet Model}
We take inspiration from the long history of using binary spins in models for neural operations~\cite{hopfield_neural_1982} and theories for their operation in neural networks (examples are discussed in Ref.~\cite{coolen_theory_2005} and references therein). Likewise, Ising spin models have been shown to be well suited to correctly predict the overall physics of ASIs because the spin's binary `up' or `down' describes the nanomagnet's north-south pole alignment. In the following, we lay out how our proposed model behaves magnetically, and then map out how that behaviour is a physical implementation of an active inference agent. In our model, an Ising spin is represented by a nanometer scale magnet shaped approximately like a rectangular needle. As single-domain nano-sized magnetic particles, each nanomagnet presents a magnetization aligned along the needle axis below a critical temperature $T_c$ without the application of an external magnetic field.  At their simplest, nanomagnets can be considered to behave like compass needles in that they attempt to align their north and south poles according to the direction of magnetic fields (but constrained by the needle geometry).  

\begin{figure}[t!] 

\includegraphics[width=0.98\textwidth]{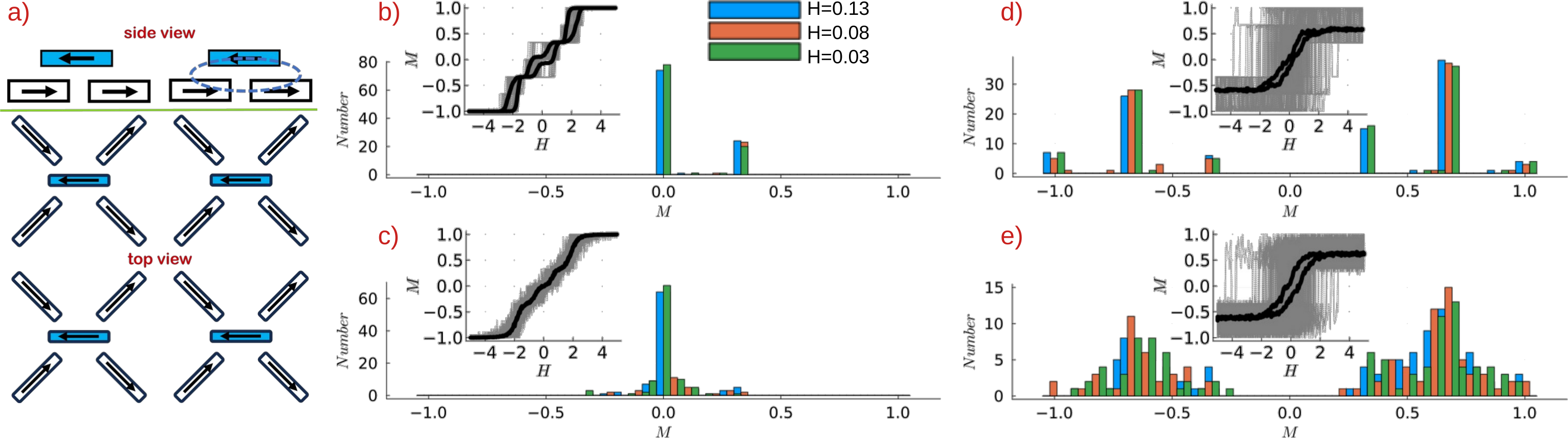}
   {\phantomsubcaption\label{fig:1a}%
	\phantomsubcaption\label{fig:1b}}
\caption{a) Geometry of the bilayer with Artificial Spin Ice for use in active inference. The top layer of spins (shaded blue in the figure) are placed above the vertices of the so-called `square' ASI geometry (unshaded). The top layer elements interact with the bottom layer through their individual (stray) magnetic fields as discussed in the text.
\\b) and c) An external field is applied directly to the hidden spin square ice layer for numerical simulation of magnetic hysteresis taken over $700$ field steps. The field is applied uniformly to all of the hidden spins and response is simulated with Monte Carlo using $N_h=100$ and $T_h=0.4$ in b) and $T_h=1.0$ in c). The states of the hidden spins are distributed as shown in the main plot at small fields for three neighbouring field steps near $H=0$ for the hysteresis shown in the insets. Only two significant peaks appear at low temperature, one at $M=0$ and the other at $M=0.4$. The hysteresis apparent in the b)$M(H)$ indicates relaxation into long lived metastable states. 
\\d and e) Hidden state response to sensory spins driven by an external field are shown in (inset to d) for $T_h=0.4$ and in (inset to e) for $T_h=1.0$. Values $T_s=1.0$ and $N_s=1$ are used in both cases. Hysteresis appears for each $T_h$ and is widest at the higher temperature. The hidden spin states occupied at small external fields are shown in the main plot (d) for the low $T_h$ case, and (e\label{fig:1e}) for the higher $T_h$ case. The wide distributions are because sensory spin response to the external field creates local fields in the hidden spin array that access configuration states which are not possible if the external field is applied directly and uniformly to the entire hidden spin array.\label{fig:1}}

\end{figure}

We propose two arrays of nanomagnets in a bilayer configuration arranged in a three dimensional geometry as sketched in figure~\ref{fig:1}a.  This structure is a modification of a geometry suggested in~\citet{begum_popy_magnetic_2022}.  The top layer is made of well-separated superparamagnetic nanomagnets whose magnetization can flip direction randomly and independently by thermal excitation as superparamagnets. However, also as with superparamagnets, the nanomagnets will try to align their magnetizations as parallel as possible to a sufficiently strong external magnetic field.  The bottom layer is an array of  nanomagnets arranged in an ASI geometry whose nanoelements are placed close enough to interact strongly with one another through their stray magnetic fields. They also respond to the stray fields produced by the top layer of superparamagnetic nanomagnets. 

Effects due to the magnetic field acting on the top superparamagnetic layer are assumed to be negligible at the bottom square ASI layer. In this sense the top layer acts as a ``sensor'' of magnetic fields from an external environment that transmits information to the ``hidden'' (from the environment fields) bottom ASI layer.  The ASI nanomagnets produce an average magnetization that can be measured and used to provide a signal which is fed back to the environment to control future input to the top-layer sensory nanomagnets.  We find that when viewed on the time-scale of the environment, the average magnetization produced by the hidden layer follows a trajectory described by the magnetic response that can converge to the neighbourhood of a preset value.  In what follows we will show that this behaviour is described accurately by the neurological free energy principle with active inference. This formalism provides a useful means of describing complex out-of-equilibrium non-linear dynamics that evolves on internal timescales that can differ from the timescale over which the environment changes.  


The bottom hidden ASI layer of figure~\ref{fig:1}a is made of $16$ spins in a so-called ``square'' ice geometry. This geometry is known to have a two-fold degenerate ground state where the two lowest energy spin configurations produce a zero average magnetization for the array.~\cite{nisoli_colloquium_2013} These ground states are characterized by ``ice rules'' that require two spins at each vertex to point inwards towards the vertex centre with the other two spins pointing out from the vertex. The stability of vertex spin configurations depends on temperature as thermal fluctuations have a finite probability to reverse spins randomly. The configurations of vertex spins can form topological excitations that propogate through the ASI and govern the emergence of global metastable nanomagnet spin configurations.~\cite{skjaervo_advances_2020}

The local fields acting on the ASI, produced by the top sensory layer, affect directly the creation and propagation of vertex excitations. We show below how these local fields manipulate relaxation dynamics in the ASI. Note that our requirements for spins in the sensory layer are that they do not reverse as easily as the spins in the ASI bottom hidden layer and are assumed to align independently of both the ASI and of one another. These requirements present challenges for practical design which are not unrelated to those addressed by magnetic data storage technologies. Some considerations are discussed more fully at the end of this paper.

\section{Properties of the Sensory Spin Layer}

We use numerical simulations to illustrate the importance of how the sensory layer facilitates sampling of hidden layer spin configurations. For the numerical simulations a $36$ element square ice is used with nine sensory control elements placed above the vertices internal to the lattice in the manner depicted in figure~\ref{fig:1} (for relevant parameters and other definitions see Methodology). For simplicity, $m_h$ is set to unity and $m_s$ then appears only in the environment field energy and the magnetic interaction between sensory and hidden spins. Monte Carlo sampling is made for an ensemble of 100 identical bilayer array replicas sampled independently.  The number of Monte Carlo Steps (MCS) performed during each time interval are specified separately for the sensory and hidden spins of numbers $N_s$ and $N_h$, respectively. The algorithm used for sampling is Glauber dynamics.

The stray magnetic field produced by a sensory spin affects most strongly the four spins that comprise the ASI vertex directly below. These additional fields on the ASI spins impacts the stability of each four-spin vertex and will bias the probability of spin reversals for each of the vertex spins. In this way the sensory layer embodies some aspects of a Markov Blanket in that it mediates information that the hidden array can receive from the environment.~\cite{friston_life_2013,kirchhoff_markov_2018,aguilera_how_2022} Information flow between the sensory and hidden spin arrays is directed in that the sensory array is designed to respond to the environment external field whereas the hidden spins respond only to the individual sensory spin fields. 

The importance of this sensory `blanket' design can be appreciated from two aspects of how spin states in the hidden layer are accessed by the external field. To illustrate this point, in figures~\ref{fig:1}b-e comparisons of state access are made between the case of an external environment field applied directly to the hidden spins without sensory spin fields and the case of only sensory spin fields acting on the hidden spins. 

We consider first the case without sensory spin mediation of the external field. In figure~\ref{fig:1b} magnetization states accessed in hidden spin array by a magnetic field applied directly to the ASI are shown for fields sampled during a hysteresis loop. The complete hysteresis loop is run with $700$ field steps as shown in the inset $M-H$ plot. In the hysteresis plot the dark line represents the average taken of $M$ over all replicas and the lightly shaded lines are the individual replica $M$ hysteresis loops. The magnetic field is applied uniformly across all spins in the hidden layer \emph{without} mediation from sensory spins. The simulation is run for $N_h = 100$ MCS which is long enough to allow the hidden spin system to relax towards long-lived meta-stable states at each applied field and the distribution of these states for three small applied fields are shown in the main plot. 

The temperature in figure~\ref{fig:1b} is $T_h=0.4$ (in units of dumbbell strength) and state occupation is shown in the main plot for the three closely spaced fields chosen near $H=0$. The distribution is discrete due to the ASI geometry which constrains possible configurational states. Only two of the possible states are occupied significantly which is consistent with the low open loop fields in the $M-H$ hysteresis. Small changes in the applied field do not excite additional states as seen by the close proximity of the three neighbouring sampled field strengths. Changes in the distribution appear at higher the temperature $T_h = 1.0$ as illustrated in figure~\ref{fig:1}c. More states are accessed at this higher temperature but the broadening is limited to states near the two accessed at lower temperature. Note also that the $M-H$ loop is closed so that the state distribution is now peaked significantly only about $H=0$.

Figures~\ref{fig:1d}d and \ref{fig:1e}e represent the case where the external environment field only acts on the sensor spin layer, and the ASI hidden layer experiences only fields generated by the sensor spins. The temperatures are $T_s=1.0$ with $T_h=0.4$ in figure~\ref{fig:1d}d and $T_h=1.0$ in figure~\ref{fig:1e}e. A small hysteresis appears in the inset $M-H$ loops, but the ASI $M$ has large fluctuations in both cases as evidenced by the replica loops. The main plots show the occupation of ASI states for (external environment) fields near $H=0$, again at three neighbouring values. The low temperature distribution has six significant peaks spread across all possible values ($\pm 1$). The spread associated with the three neighbouring $H$ values is broader than what occurred for the directly applied field case shown in figure~\ref{fig:1b}. The higher temperature example in figure~\ref{fig:1e}e shows a much broader distribution of accessed states as well as a broader loop in the hysteresis plot inset.


\begin{figure}[t!] 

\includegraphics[width=0.98\textwidth]{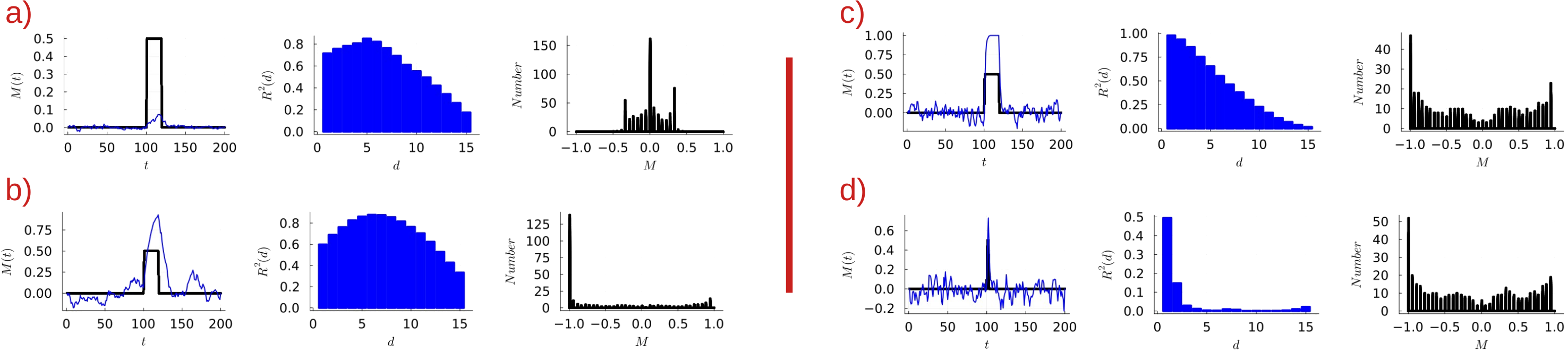}

\caption{Pulse field profile and averaged $M$ response are shown in the left panels. Cross-correlation is shown in the middle panels, and the distribution of hidden spin states is shown in the rightmost panels. In (a\label{fig:2a}) a $20$ time step wide pulse is applied as directly without sensory spin mediation. In (b\label{fig:2b}) the pulse is delivered to the hidden spin layer via the sensor layer. The sensory spins are sampled at $T_s=0.1$ with $N_s=1$. In (c\label{fig:2c}) the 20 time step pulse width is delivered to the hidden spin layer through the sensory spin layer with $T_s=0.1$ and $N_s=10$. In (d\label{fig:2d}) $T_s=0.1$, $N_s=10$ and the pulse width is $4$.\label{fig:2}}

\end{figure} 


The differences between how the fields on the hidden layer are generated shown in  figure~\ref{fig:1} can be understood in terms the spin flip processes responsible for changing $M$ values. The nanomagnet elements in the hidden layer interact strongly, and alignment of these elements occurs  generally in square ASI via a type of avalanche process. When the external field is applied directly to all the square ASI elements, excitation of these processes will most likely begin at array edges where the total local interaction field acting on nanomagnets is weakest. Avalanches for an ASI directly responding to a uniform applied field will typically begin only at the edges.  When the sensor layer is present, sensor generated local fields can destabilize orientations within the hidden layer ASI at sites away from the array edges. This leads to a greater probability of nanomagnet reversal avalanches nucleated at sites within the ASI array closest to the sensor spins.

From these examples, one can see that the top sensor layer provides local fields that facilitate broader sampling of states than a globally applied field can. This is further enhanced by spatial differences of local fields that will appear because each sensory spin fluctuates randomly due to thermally driven reversal but in a manner that the average remains consistent with the value of the external field. The ability to leverage multiple states through defects or direct control of ASI states was noted several years ago by Budrikis using graph theoretical methods\cite{budrikis_diversity_2011,budrikis_disorder_2012}. The behaviour observed in the present paper can be understood similarly in that local fields acting on a subset of spins can open pathways for avalanche processes that would otherwise have a low probability of occurring. This gives the overall hidden layer ASI access to a larger portion of configurational phase space through the top sensor layer than is possible when a field is applied directly to the ASI.  Another perspective is to consider that the spread of states sampled through the ensemble of hidden spins is wide because there exists a multitude of possible trajectories available to the correlated hidden spins as they evolve in time towards a lower energy configuration. This distinguishes the hidden spin system from a purely random Markov sampling of states. A range of possible trajectories towards some minimal energy is fundamental to the operation of the system in the active inference applications that will be discussed later. 

The temporal duration of a signal (i.e. the time-span or pulse-length of the applied field of the environment) is also important, and here too the bottom hidden layer's response is facilitated by the top sensor layer.  To understand this, we  examine the bottom hidden layer response to a pulsed environment field without a top sensor layer.  Figure~\ref{fig:2a}a contains results for a pulsed external field where $H$ lasts for 20 time steps (with no sensory spin mediation) at $T_h = 1.0$. The left panel of figure~\ref{fig:2a}a shows $H$ for a 20 time-step wide pulse profile (black line), while the $M_t$ response averaged over all replicas of the system is shown in blue. The middle panel of figure.~\ref{fig:2a}a is the cross-correlation response between the pulse and the $M_t$ response to the pulse for delay times ranging up to 15 time steps. The correlation is normalized and calculated as described in~\citet{hon_numerical_2021} using the definition
\begin{equation}
	R^2(d)=\left[ \sum_t \frac{(M_t-\langle M_t \rangle_t)\{p_t(d)-\langle p_{t-d} \rangle_t\}}{(N-1)\sigma_M\sigma_p}\right]^2 
	\label{eq:crosscorr}.
\end{equation}
The averages at each time step are taken over the replica-averaged hidden layer magnetization response $M_t$ calculated at each time step $t$, and the pulse input to the sensory spins $p_t$. The averages are over the time interval sampled, and $\sigma_M$ and $\sigma_p$ are the corresponding standard deviations. The delay between $M_t$ and $p_t$ is $d$.  The cross-correlation presented in figure~\ref{fig:2a}a reaches a maximum at around six time steps into the pulse, and decays rapidly thereafter. The states accessed throughout all times  are tightly grouped around $M_t=0$ as shown in the right panel of figure~\ref{fig:2a}a.

The effects of sensor layer mediation are shown in figure~\ref{fig:2b}b where $T_s=0.1$ and $N_s=1$.  Clearly, sensor-layer-mediated input to the hidden spins increases dramatically the hidden layer's sensitivity and response to the field pulse.  The correlation $R^2$ and distribution of states (middle and right panels of figure~\ref{fig:2b}b, respectively) present a delayed, but significant, $M(t)$ response to the onset of the pulse, with a clear asymmetry for $M=-1$ in the state distribution that corresponds to the pulse.

The results of a larger sensory spin MCS, with $N_s=10$, is shown in figure~\ref{fig:2c}c with $T_s=0.1$.  The $M_t$ response is now better synchronized with the  pulse as can be seen from the response profile and the $R^2$ correlation. The distribution of states has states spread away from the $M_t=-1$ value.  Lastly, figure~\ref{fig:2d}d illustrates sensitivity to pulses through sensor layer mediation. Here the field pulse width is reduced from 20 to 4 time steps.  We find that when there is no sensor layer, the $M(t)$ response of the hidden layer to the significantly shorter field pulse is very weak (not shown).  The left panel of figure~\ref{fig:2d}d shows that the hidden layer $M$ response to the shorter field pulse is substantial when done through the sensory spins local fields (right panel). Also, strong correlation with the pulse remains, and the distribution of states is similar to that in figure~\ref{fig:2c}c.

This sensitivity to changes in environment external field and the ability to activate a range of states is perhaps central to the operation of the system for active inference. As discussed earlier, the sensor layer creates local fields that, through avalanche dynamics, instigate configurational changes that drive the hidden system through its spin state phase space. In what follows, it will be seen that this sensitivity enables the system to search widely for configurations that, when active inference is enabled, direct the hidden spin evolution trajectory towards targets.

\section{Sampling the Variational Free Energy and Implementation of Active Inference}

A few definitions are required before we can describe how the above bilayer system is able to perform active inference. The theory we describe is a formulation in terms of stochastic differential equations proposed by Friston~\cite{friston_free-energy_2007}. Only a summary of the essential points is presented here as the complete theory is well described in numerous other papers.  The formulation and examples we use largely follow the description presented by~\citet{buckley_free_2017}. 

The external environment defines a time sequence that is sampled regularly by the sensory spins. The variables sampled consist of data encoded as time varying fields and can represent different components such as positions, velocities and accelerations, and other rates of change. In the formalism used here, variables are defined as generalized coordinates which facilitate the multiple timescale feature highlighted earlier and enables definition of continuous time evolution of otherwise stochastic quantities. A generalized $x(t)$ coordianate is defined by $x(t)$ and its derivatives. These can be represented as a list $\tilde{x}$ of data values where each component is a derivative with respect to $t$. To simplify the notation, a component of $\tilde{x}$ is denoted by $x_\alpha=d^\alpha x/dt^\alpha$.  A notable feature of this representation is that the action of a derivative on $\tilde{x}$ is defined as the promotion of a component to its next higher order derivative component. This operation is denoted by the operator $D$ where $D x_\alpha \rightarrow x_{\alpha+1}$ (within the same vector component set $\tilde{x}$). 
		
The configurational states of the sensory spins create a mapping of the environment with a component of $\tilde{\phi}$ corresponding to a component of $\tilde{x}$ at time $t$.  Here, a sensory state $\phi_\alpha$ is sampled during a MCS over the sensory spins in the presence of an external environment magnetic field, representing a component of $\tilde{x}$.  The value assigned to $\phi$ is an average over $N_s$ MCS for the ensemble of system replicas. An average generated this way is made for each member of $\tilde{x}$, and thereby mapped to the corresponding values of $\tilde{\phi}$.  
				
In a similar manner, state averages $\tilde{\mu}$ are defined for the hidden spins where an average of an ensemble of hidden spin states is mapped to the $\tilde{\phi}$ (which are in turn mapped to the $\tilde{x}$). Note that the timescale for this mapping is not specified by a particular $t$, but depends on the number of MCS taken during a particular $t$. This provides an interesting separation of time scales between the changing values of the $\tilde{x}$ that occur on the environment time, and the response of the bilayer components $\tilde{\mu}$ and $\tilde{\phi}$ that are changing during stochastic thermal relaxation processes.

The probability for a thermal reversal of a nanomagnet element  `spin' during a time interval $\Delta t$ depends on the energy difference between an initial and possible final state and the attempt frequency. This can be described by an Arrhenius relation, $\nu \exp(-\beta\epsilon)$ where $\beta$ is the inverse temperature and $\epsilon$ the energy difference between states.  The attempt frequency $\nu$ measures the number of pathways to an energy barrier saddle point relative to the number of pathways over that barrier. The Monte Carlo algorithm used herein does not take $\nu$ into account directly.  However the number of MCS determine to what degree configurations sampled during one $\Delta t$ time interval are correlated with those of the previous interval. This provides a very rough analogy to the prefactor $\nu$, and captures some sense of the different timescales for relaxation that the spins in the two layers experience.  Larger MCS values correspond to less `memory' between time steps (determined by the environment applied field). The parameters $N_s$ and $N_h$ thereby determine the rates at which the spins in the layers remain coherent over time with respect to the `environment clock' determined by $\Delta t$. As will be seen below, judicious choice of $N_s$ and $N_h$ are important to optimize performance of the bilayer system for active inference.

During thermal relaxation, transitions change states $\tilde{\mu}$ to new average values. A variational Bayes relaxation is used to infer the best estimate of the distribution describing the joint probability $P(\mu,\phi)$ of having a state $\mu$ when the sensory spins are in a state $\phi$. The algorithm for this describes the evolution of a trajectory for each component of $\tilde{\mu}$ defined by

\begin{equation}
	\frac{d\mu_\alpha}{dt} - D\mu_\alpha = -\kappa \frac{\partial E( \tilde{\mu},\tilde{\phi} )}{ \partial \mu_\alpha } .  
	\label{eq:traj}
\end{equation} 

\noindent The quantity $E( \tilde{\mu},\tilde{\phi} )$ plays the role of a negative entropy as defined in ensemble learning\cite{buckley_free_2017} which assigns $E$ to a distribution $P(\mu,\phi)\sim \exp\{-\beta E( \tilde{\mu},\tilde{\phi} ) \}$
 that represents the unknown joint probability relating hidden states to sensory states. 

As key point of the present paper, $E( \tilde{\mu},\tilde{\phi} )$ is here viewed as a thermodynamic energy associated with the bilayer system. The gradient in Eqn.~\ref{eq:traj} suggests an interpretation of the right hand side as a conjugate field to $\mu_\alpha$ driving states of the hidden spins along a trajectory that leads to the condition $d\mu_\alpha/dt - D\mu_\alpha =0$. This condition describes a trajectory through state phase space where the time evolution of the average $\tilde{\mu}$ is directed towards the maximum of the distribution that is being inferred\cite{balaji_continuous-discrete_2009}. 

These ideas are illustrated for the ASI bilayer through the results presented below. The first example is for tracking of the magnetization $M$ to a target parameter value subject to unknown (to the bilayer spins) external constraints imposed by the environment. 

This example is set up with a model adapted from~\citet{baltieri_active_2019}. Here, the environment is defined by a one dimensional equation of motion specified as
\begin{equation}
	\frac{dx(t)}{dt} = - \gamma x(t) + a(t).
	\label{eq:tracking}
\end{equation} 
This equation describes dissipative motion along $x$ that would decay to zero via a friction $\gamma$ unless offset by a velocity term $a(t)$. The target specifies a fixed velocity $dx/dt = v_d$ the system should arrive to  as it relaxes to the condition $d\mu_\alpha/dt - D\mu_\alpha =0$. 

The sensory spins' response to the environment field are mapped to position and velocity information through generalized coordinate vectors $\tilde{x}=\{{x(t),dx(t)/dt}\}$ at regular time steps~\cite{balaji_continuous-discrete_2009,balaji_bayesian_2011}. An important note is that here the same bilayer is used for each component of $\tilde{x}$ with each component updated sequentially as discussed below. Each of the bilayer's hidden spins will respond with an overall magnetization $M_\alpha$ that is defined by an ensemble average over a quantity $\chi(B_\alpha)$ via
\begin{equation}
  	M_\alpha = \langle\chi(B_\alpha)\rangle_T
   	\label{eq:chi},
\end{equation}
where $\chi$ is the magnetization measured along one direction of the spin lattice array that experiences the field $B_\alpha$.  The average $M_\alpha$ replaces $\partial E( \tilde{\mu},\tilde{\phi} )/\partial \mu_\alpha$ in Eqn.~\ref{eq:traj}.

The quantity $B_\alpha$ is a field corresponding to a linear combinations of $\tilde{\mu_\alpha}$ and $\tilde{\phi_\alpha}$ components that are determined in the following way:  $\chi(B_\alpha)$ samples a component of the log joint probability $P(\tilde{\mu},\tilde{\phi})$. Following~\citet{friston_variational_2007}, under the assumption of statistical independence,  one arrives at the factorizations
\begin{equation}
	P(\tilde{\mu}|\tilde{\phi}) = \prod_\alpha{ P(\mu_\alpha | \phi_\alpha) }, 
\label{eq:decomp1}
\end{equation}
and
\begin{equation}
P(\tilde{\mu}) = \prod_\alpha { P(\mu_{\alpha+1} | \mu_\alpha) }
\label{eq:decomp2}
\end{equation}
Sampling distributions produced by the log of these probabilities then reduces the problem to a summation of $M_\alpha$ terms where the $B_\alpha$ in Eqn.~\ref{eq:chi} are determined by the conditional variables appearing in the $P(\tilde{\mu}|\tilde{\phi})$. For example, the ensemble average of $\log P(\mu_\alpha | \phi_\alpha)$ corresponds to evaluating $\langle\chi(\phi_\alpha-\mu_\alpha)\rangle_T$. Note that when using the Laplace approximation of Gaussian distributions for each $P$, the precision of the Gaussians enter as adjustable parameters that can optimize the ability of the system to relax towards target values of sensory input. In the bilayer, the inverse temperatures $\beta_s$ and $\beta_h$ become the corresponding adjustable parameters, although $N_s$ and $N_h$ also play a role.

The final step is to include active inference. Active inference provides a feedback from the hidden spins to the world environment, and in the present context is represented by the $a(t)$ appearing in Eqn.~\ref{eq:tracking}. This term is defined by imposing a functional dependence for $\phi(a)$ such that it appears as a time dependent constraint on minimization to the steady state condition of Eqn.~\ref{eq:traj}. A time evolution defined as a force
\begin{equation}
	\frac{da(t)}{dt} = - \kappa \frac{d\phi}{da} \frac{dE( \tilde{\mu},\tilde{\phi} )}{d\phi}
	\label{eq:action}
\end{equation}
is assumed. Proceeding as above, the bilayer equivalent replaces $E$ with $\langle\chi(B_\alpha)\rangle_T$ to determine the form of Eqn.~\ref{eq:action}. The manner in which $a(t)$ enters $\tilde{\mu}$ is through definition of a target probability $P_S$. In general this corresponds to assigning $a(t)$ to a particular component (or components) of $\tilde{\mu}$. 

In this particular tracking example, the factorization requires only two generalized components for each of the sensory and hidden spin states. They are $\phi_0$ and $\phi_1$ for the spin layer, and $\mu_0$ and $\mu_1$ for the hidden layer.  The target appears in the $\mu_1$ component via the transition probability $P_S(\mu_1| v_d-\alpha\mu_0)$. Note that the factorization of $P(\tilde{\mu})$ is truncated by requiring $\mu_2 = 0$.

\begin{figure}[t!] 

\includegraphics[width=1.00\textwidth]{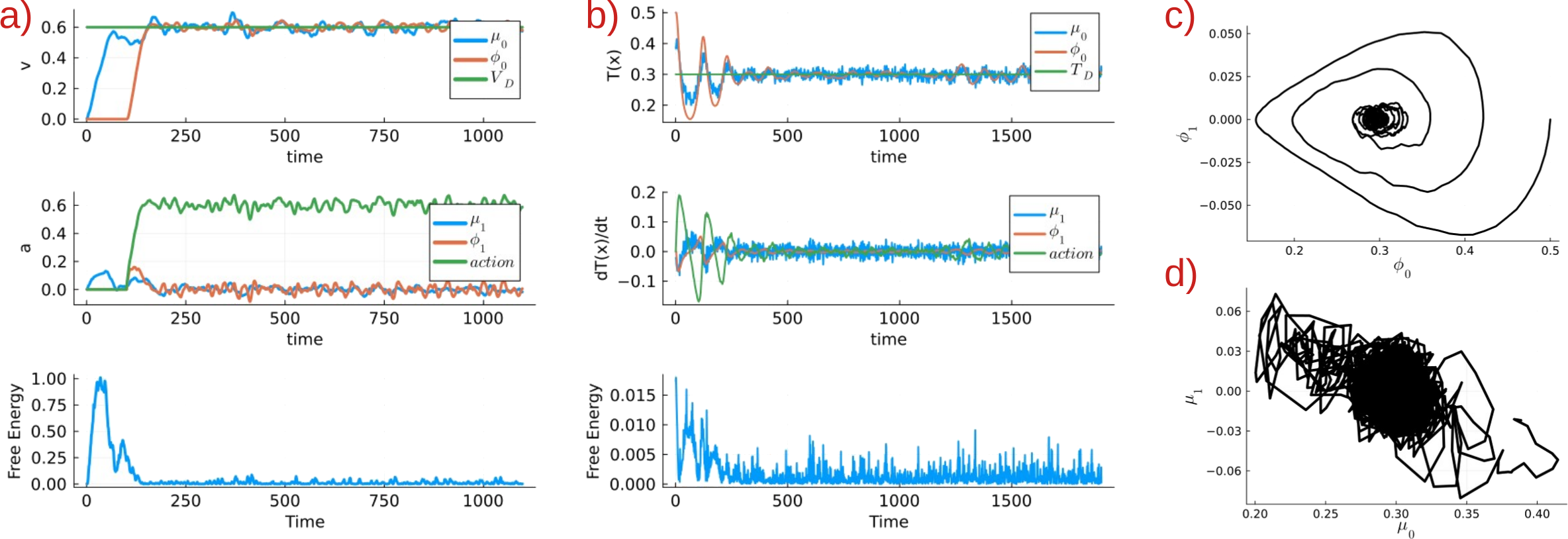}

\caption{ a) and b) present two example applications of active inference and the variational free energy principle. In (a\label{fig:3a}), sensory layer spins receive environment position and velocity information and active inference generated by hidden layer spin states drive the system towards a target velocity $v_d=0.6$. In (b\label{fig:3b}) environment measures are determined by a function $T(x)$ and its derivative $dT(x)/dx$, but the sensory spins receive only the values $T$ and $dT/dx$ without information of $x$ or the functional form of $T(x)$. Active inference drives the system to a specified target $T(x) = T_d$.  The trajectory of the sensory spin states is shown in (c\label{fig:3c}) for the example in (b). Active inference feeds information derived from the hidden spin state averages to the environment that causes changes in subsequent sensory input. In this way the trajectory of states within the hidden layer are directed toward the target goal by the sensory spins in response to changes in the environment. The corresponding trajectory for the hidden spins is shown in (d\label{fig:3d}).
}
\label{fig:3}
\end{figure}

The resulting evolution equations for the first example shown in figure~\ref{fig:3a} are
\begin{equation}
	\frac{d\mu_0}{dt} = \mu_1 + \langle\chi({ \beta_z (\phi_0-\mu_0))\rangle_T - \alpha \langle\chi(\beta_w (\mu_1 + \alpha\mu_0-v_d) })\rangle_T, 
		\label{eq:eomtrack1}
\end{equation}
\begin{equation}
	\frac{d\mu_1}{dt} = \langle\chi(\beta_z (\phi_1-\mu_1))\rangle_T - \langle\chi(\beta_w (\mu_1 + \alpha\mu_0-v_d))\rangle_T, 
		\label{eq:eomtrack2}
\end{equation}
\begin{equation}
	\frac{da}{dt} = -\langle\chi(\beta_z (\phi_1 - \mu_1))\rangle_T.
	\label{eq:eomtrack3}
\end{equation}


In this expression, a generalization of the effective temperature is introduced which allows the temperature of the hidden layer for each bilayer to be treated as a parameter. In~\citet{baltieri_active_2019} separate precisions are defined for the Gaussian distributions that are instead here sampled from the ASI. Here we define corresponding parameters as $\beta_z=\beta_h\Pi_z$ and $\beta_w=\beta_h\Pi_w$ with additional parameters $\Pi_z$ and $\Pi_w$ that are used to adjust the probabilities affecting the $\phi_\alpha$ and $\mu_\alpha$ terms. This generalization provides a functionality analogous to the precisions used in~\citet{baltieri_active_2019}. Unless specified otherwise, for simplicity we specify only $T_h$ (so that $\beta_z = \beta_w$) or $\beta_z$ and $\beta_w$ separately. We also note that $T_s=0.5$ for all simulations that follow.  

In order to confirm minimization, a measure of the free energy is defined as the lowest order contribution to the factorized $\log P(\tilde{\mu},\tilde{\phi})$:
\begin{equation}
	F \approx \frac{1}{2} [{\beta_z (\phi_0-\mu_0)^2 + \beta_z(\phi_1-\mu_1)^2  + \beta_w (\mu_1 + \alpha\mu_0-v_d)^2 }].
	\label{eq:fetrack}
\end{equation}
The panels in figure~\ref{fig:3a}a are the result of iterating Eqs.~\ref{eq:eomtrack1} through~\ref{eq:eomtrack3}. The top panel of figure~\ref{fig:3a}a shows how $\mu_0$ and $\phi_0$ evolve in time with reference to the target $v_d$. The middle panel shows the time evolution of $\mu_1$, $\phi_1$ and $a$. The corresponding free energy is in the bottom panel. The system evolves to the target value of $v_d=0.6$ using $\beta_z = 10$ and $\beta_w=1$. Here $N_s=1$, and $N_h=1$ at each time step. This system still arrives at the target for larger values of $N_s$ or $N_h$ with generally less noise. The ability of the system to reach the target is sensitive to the values of the inverse temperatures in analogy to sensitivity to precisions noted in~\citet{baltieri_active_2019}.  

In the above example a model of the environment was encoded in the hidden spin system through $P_S$ by inclusion of the $\alpha$ dissipation term. This next example illustrates that only the target need be encoded in the hidden states, and additional information can be inferred from the environment at the sensory level.  The basis for this example is the thermostat problem presented in Buckley~\cite{buckley_free_2017} and discussed in the context of proportional integral derivative (PID) control in~\citet{baltieri_active_2019,baltieri_bayesian_2020}. The sensory input is from two functions of position $x$ defined by
\begin{equation}
	T(x) = \frac{1}{1 + x^2},
	\label{eq:thermoT}
\end{equation}
and 
\begin{equation}
	\frac{dT(x)}{dx} = -\frac{2x}{(1 + x^2)^2}.
	\label{eq:thermoDT}
\end{equation}
Action is included in $dx/dt$:  
\begin{equation}
	\frac{dx(t)}{dt} = a(t)
	\label{eq:kthermox},
\end{equation}
In this model, target information $T_d$ is included through $P_S(\mu_1 | \mu_0 -T_d )$, but no information about $\tilde{x}$ is passed to the bilayer. The equations of motion are truncated by requiring $\mu_3=0$. The equations for $\tilde{\mu}$ are of the same form as in~\citet{buckley_free_2017}, repeated here for completeness:
\begin{equation}
	\frac{d\mu_0}{dt} = \mu_1 + \langle\chi({ \beta_z (\phi_0-\mu_0))\rangle_T - \langle\chi(\beta_w (\mu_0 + \mu_1 - T_d) })\rangle_T, 
	\label{eq:eomthermo1}
\end{equation}
\begin{equation}
	\frac{d\mu_1}{dt} = \mu_2 + \langle\chi({ \beta_z (\phi_1-\mu_1))\rangle_T - \langle\chi(\beta_w (\mu_0 + \mu_1 - T_d))\rangle_T + \langle\chi(\beta_w(\mu_1+\mu_2) })\rangle_T, 
	\label{eq:eomthermo2}
\end{equation}
\begin{equation}
	\frac{d\mu_2}{dt} = -\langle\chi(\beta_w(\mu_1+\mu_2))\rangle_T.
	\label{eq:eomthermo3}
\end{equation}
The time dependence of the action for this example given by
\begin{equation}
	\frac{da}{dt} = -\langle\chi(\beta_z(\phi_1-\mu_1))\rangle_T\frac{dT(x)}{dx}.
	\label{eq:eomthermo4}
\end{equation}

\noindent Results are shown in figure~\ref{fig:3b}b for the same temperatures as above but with $N_s=10$ ($N_h$ remains $1$). The increased value of $N_s$ helps optimize the trajectory to arrive at the target value $T_d=0.3$. The $\mu_0$ and $\phi_0$ time evolutions are shown in the top panel, $\mu_1$, $\phi_1$ and $a$ are in the middle panel, and the free energy is in the bottom panel; it is clearly minimized as the system evolves. The decaying oscillations observed in the $\tilde{\mu_0}$ and $\tilde{\phi_0}$ are strikingly reminiscent of PID control behaviour as discussed by~\citet{baltieri_bayesian_2020}. The authors therein note that optimization of parameters can be performed to enhance the decay toward the target, and we find the same.

It is illuminating to track the evolution through $\phi_0$ and $\phi_1$ when viewed as a phase space trajectory. The trajectory corresponding to the evolutions of the sensory input shown in figure~\ref{fig:3c}c is for the time evolution of hidden states shown in figure~\ref{fig:3d}d. The corresponding trajectory for the hidden state $\tilde{\mu}$ is quite different as shown in figure ~\ref{fig:3d}d. The sensory inputs circle through $\tilde{\phi}$ to oscillate around the target value. The hidden states $\tilde{\mu}$ lie roughly along a quasi-linear line centred about the neighbourhood of the target value. Gradients of the free energy direct evolution of the hidden states towards a most likely maximum that lies along this line at the target value. There is a strong dependence on the number of sensory spins, $N_s$, where the system has difficulty finding the target for small values causing it instead to relax to $\phi_0=0$.   The optimal parameter values for achieving the target also depend on the magnitude of the target $T_d$. As noted earlier, thermal evolution during environment time scales within the hidden spin system depend on temperatures and MCS. 

Details of how these parameters affect the trajectories and relaxation are not well understood at present, but some qualitative observations can be reported. The results shown in figure~\ref{fig:3} require an amount of environmental time to relax towards the target, and also depend on $T_h$ and $N_h$. Increasing $N_h$ generally appears to increase the correlation between $\tilde{\mu}$ and $\tilde{\phi}$, while introducing more noise into $\tilde{\mu}$ and the free energy. $T_h$ (and its generalization to $\beta_z$ and $\beta_w$) is an important optimization parameter for achieving the lowest free energy for different $T_d$. 

\begin{figure}[t!]  

\includegraphics[width=0.98\textwidth]{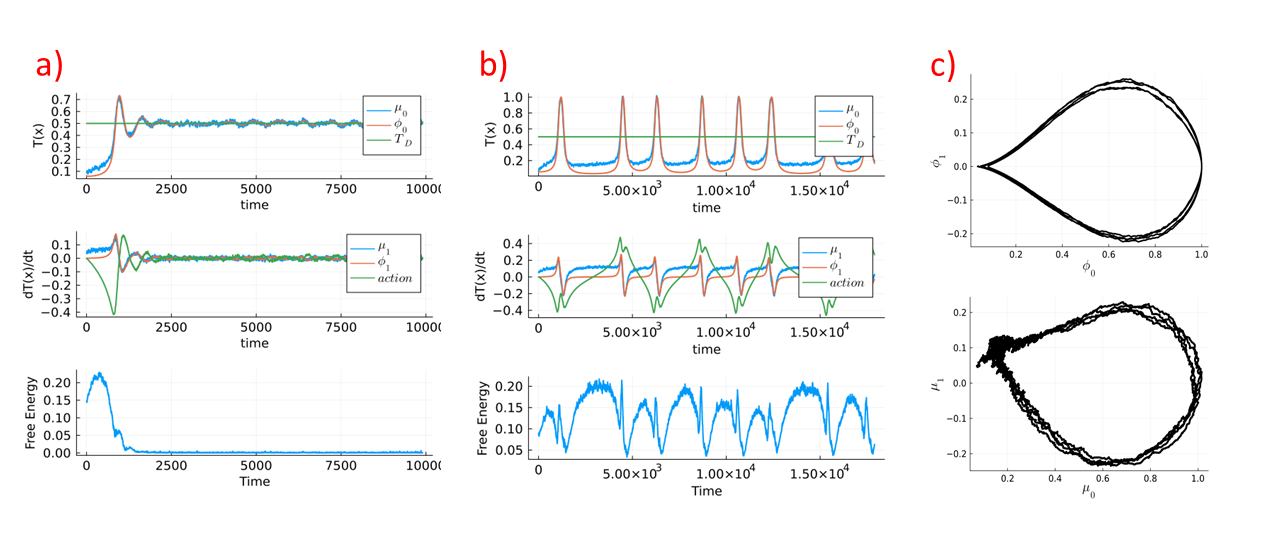}

\caption{The oscillations of $\tilde{\mu}$ and $\tilde{\phi}$ relax toward a target $T_d=0.5$ in (a\label{fig:4a}) for parameters $\beta_z=3.0$ and $\beta_w=0.75$. Similar oscillations are found using the same parameters when the $\langle\chi\rangle_T$ is replaced with a Gaussian distribution approixmation. A very different quasi-periodic oscillation arising from the ASI through  $\langle\chi\rangle_T$ is found for $\beta_z=2.0$ and $\beta_w=0.5$ as shown in (b\label{fig:4b}). A Gaussian approximation for these $\beta_\alpha$ values is instead similar to the results shown in (a). The oscillations correspond to limit cycles as seen with the trajectories for the corresponding $\tilde{x}$ and $\tilde{\phi}$ displayed in  (c\label{fig:4c}).\label{fig:4}}
	
\end{figure}

A very interesting aspect is the dependence on $\beta_z$ and $\beta_w$ of the time evolution for $\tilde{\phi}$ and $\tilde{\mu}$. Unusual oscillations can appear for different temperatures and targets that may contain information about internal state selection within the hidden layer. These are aspects currently under study that are outside the scope of the present work, but an example is shown in figure~\ref{fig:4}. 

Results for a $T_d=0.5$ with $\beta_z=3.0$ and $\beta_w=0.75$ are shown in figure~\ref{fig:4a}a. These are qualitatively similar to those shown in~\ref{fig:3b}b although with noticeable differences that illustrate sensitivity of the nanomagnet dynamics to temperature. We note that the results can be fit using hyperparameters of a Gaussian distribution approximation for the $\tilde{\mu}$ distributions (which is done by replacing the $\langle\chi\rangle_T$ with Gaussians as in~\citet{buckley_free_2017}). Using instead $\beta_z=2.0$ and $\beta_w=0.5$ we find the results shown in figure~\ref{fig:4b}b. Quasiperidic oscillations in $\tilde{\phi}$ and $\tilde{\mu}$ which we were not able to reproduce using the Gaussian approximation. The corresponding trajectories for the sensory and hidden averages are shown in figure~\ref{fig:4c}c. The oscillations appear similarly for longer time integrations suggesting a stable limit cycle. In any case, this strong dependence on temperature and precision weightings suggest that non-trivial dynamics can arise under certain circumstances. Moreover, there appears to be sensitivity to details of the ASI design as we find different behaviour when we change the lattice geometry from square to pinwheel by rotating the nanoelements.  


\section{Conclusions}

Perhaps the most interesting aspect demonstrated in this paper is the possibility of studying transition dynamics in experimentally accessible magnetic and nonmagnetic systems using active inference. Viewed as a methodology for probing system dynamics, the mathematical treatment and intrinsic separation of timescales underlying the variational free energy and active inference theory may inspire new ways of studying experimentally a range of complex systems. There are few reports in the literature of applications of the neurological free energy principle in physical experiments. One example is saccadic eye motion\cite{adams_active_2015,adams_dynamic_2016} and another is a recent report on small assemblies of neurons\cite{isomura_experimental_2023}. Trajectories of quantities defined for this paper ($\tilde{\mu}$ and $\tilde{\phi}$) are sensitive to parameters of the theory that can be associated with \emph{measurable} properties of the system under study. Our purpose has been to suggest a relatively simple, non-biological system that can serve as a testing ground for exploring this approach to active inference  theory in contexts that can be studied experimentally.
 
The essential component of our approach is the sensor layer and how it provides an interface to the measured environment in a manner analogous to a Markov Blanket (as discussed by Kirchhoff et al.\cite{kirchhoff_markov_2018}). The mechanisms at play in the bilayer system involve local fields generated by the sensory spins in response to input from an environment. These local fields act on the spins in the hidden layer, enabling expedited relaxation towards a minimum free energy defined by a target constraint imposed through active inference. The ability to sample regions of hidden spin state space in this way is reminiscent of, but quite distinct from, other strategies sometimes used for escaping long lived metastable states in computations, such as parallel tempering\cite{banos_nature_2010}. It is interesting to speculate that the need to access a range of states relevant for achieving a target may be analogous to the requirement of sufficient memory and computational capacity for a system to be suitable for reservoir computing~\cite{hon_numerical_2021}. 

It is also interesting that the trajectory through states of the hidden spins that satisfies Eqn.~\ref{eq:traj} can be described as a temporally shifted reference frame under the Fokker-Planck Kolmogorov formalism.\cite{koudahl_worked_2020,balaji_continuous-discrete_2009,friston_free_2023} From this perspective, the condition for a stationary trajectory, $d\mu_\alpha/dt - D\mu_\alpha =0$, matches the environment time scale governing input to the sensory spins with the time scale over which internal hidden state processes are thermally averaged by state transitions. As noted by Balaji et al.\cite{balaji_bayesian_2011}, the stationary trajectory condition means the hidden spin states evolve along time dependent free energy gradients that lead toward target values. In the simulations discussed above, this condition assigns a physical time interval to thermal transition probabilities sampled using Glauber dynamics. It may be possible to exploit this feature to probe timescales for state transitions processes in experiments with systems whose complex energy landscape are being measured, such as spin- or structural glasses. We note that spin glasses have been explored somewhat in regards to active inference from a different direction by~\citet{heins_spin_2023}.  

The spin models we used are simple yet appear sufficient to capture the essential features displayed in experiments using actual nanomagnets. Some implementation of active inference may be possible to observe with suitably designed nanomagnet arrays using fabrication technology that is already available. For simplicity, we have here considered only magnetic interactions from the external applied field and from the nanomagnets in the bilayer. However our model creates practical challenges for implementation because of our requirement that the environment magnetic field affect directly only the top layer of magnetic nanoelements. There are a number of different strategies that would meet this requirement. For example, one could use vertical magnetic sensor elements with horizontal environment fields or variations of techniques for domain switching as used in magnetic data storage. Alternatively, a voltage actuated sensor layer using multiferroic nanomagnets such as $\epsilon$-Fe$_2$O$_3$\cite{Nickel.2023} would create local magnetic fields. Many of these technologies also permit integration into other spintronic and CMOS-based devices.  

Concerning possible applications, a great potential of creating energy efficient platforms for machine learning algorithms using nanomagnetic arrays have been identified and some in cases demonstrated in ASI\cite{jensen_reservoir_2020,hon_numerical_2021} as well as other magnetic material platforms.\cite{torrejon_neuromorphic_2017-1,grollier_neuromorphic_2020} Most recently, magnetic nanoparticle artificial spin ice configurations have been studied experimentally\cite{gartside_reconfigurable_2022} whose state configurations are manipulated locally and detected using microwaves. The energy efficiency of nanomagnetic systems for machine learning applications, as well as the microwave properties that can be associated with configurational states, make these systems attractive for practical use. The ability to incorporate active inference into architectures based on nanomagnets would be a new aspect to pursue. A very interesting direction to explore would be implementation of hierarchical structures\cite{lee_hierarchical_2003} where multiple sensory and hidden layers could be connected, and features associated with learning and other complex tasks explored. Although not reported here, stacked bilayers of the form presented above appear able to mimic some features of learning at different timescales that can facilitate optimizations toward target values.  

Finally, the nanomagnet architecture used here is only one example ASI geometry. Although not reported here, some  variants of the square ASI geometry also display analogous properties under active inference. An advantage to using ASI as a model platform is the freedom to impose constraints and introduce frustration that can be studied as pathways through configurational states. These can be analysed in a variety of ways including graph theoretical methods as discussed for athermal ASI by~\citet{BUDRIKIS2014109}. Moreover,  pathways through state configurations in these systems can often be understood via topological excitations with rules determined by allowed transitions. Active inference methodologies may be helpful in this context to understand how trajectories through configuration space can be manipulated via environmental control of topological excitations.  For this reason it would be particularly exciting to examine in non-Ising spin systems where topological excitations arise through competing interactions. A prime example is magnetic skyrmions in thin film geometries which enable control and detection via local electric potentials. Skyrmions are topologically protected spin textures and display non-linear response to applied fields and undergo transitions through mediating metastable states.~\cite{desplat_thermal_2018,desplat_paths_2019} These systems can display a sufficiently complex state space for applications in reservoir computing~\cite{PhysRevApplied.14.054020,raab_brownian_2022}.

\section{Methods}

Interactions in our numerical model of the bilayer system are provided by magnetic fields produced by the nanomagnets.  The geometry used for the hidden spins is defined by a square lattice of lattice constant $a$ with elements of length $\ell=a/2$ aligned diagonally within each unit cell. The fields produced by the sensory spins have strengths in the hidden spin layer that depend on the layer separation distance. The separation assumed throughout is $a/10$.  Sensory spin nanomagnet elements have a magnetic `spin' moment $m_s$ while ASI hidden elements have a $m_h$ `spin' moment for each element. Each nanomagnet spin is assumed to be independent of temperature for the ranges considered here.  Nanomagnet generated fields acting on a single nanomagnet are represented as a sum over all other nanomagnet spins at distances $r_{ij}$ with directions specified by unit vectors for each spin $\hat{\epsilon}_i, \hat{\epsilon}_j$. The corresponding interaction energy in the dipole limit is given by
\begin{equation}
D= \mathcal{D}  \sum_{i,j} \left[  \frac{ \hat{\epsilon}_i\cdot\hat{\epsilon_j} }{|r_{ij}|^3 } -3 \frac{ (\hat{\epsilon}_i\cdot\vec{r}_{ij}))(\hat{\epsilon}_j\cdot\vec{r}_{ij})) }{|r_{ij}|^5} \right]
\label{eq:dipole}
\end{equation}

\noindent The prefactor contains the nanomagnet magnetization and volume as $\mathcal{D}=\mu_0 /4\pi a^3$ .  In our model all dipole sums are approximated by dumbbell charges as outlined in~\citet{castelnovo_magnetic_2008} which takes into account the length of the nanomagnet elements.

The sensor layer nanomagnets' alignment with the external field is stochastic in that each nanomagnet's spin has a finite probability of aligning with the external field. In this regard, the sensory and hidden layer spins experience different temperatures $T_s$ and $T_h$ due to the different magnetic properties of each layer. The probability of a magnetization reversal for a nanomagnet at a temperature in a field specific to the layer is described by an effective temperature.  Spins in the sensory layer are at an effective temperature ${\beta_s}^{-1} = k_B T_s/m_s $ while the effective temperature of the hidden spins is ${\beta_h}^{-1} = k_B T_h/m_h $, where $k_B$ is Boltzmann's constant. 

The energy used for Monte Carlo simulations includes field and dipole interaction terms acting at each spin, and is used to determine the energy change $\Delta E$ for a possible spin reversal at each site. With a spin $\sigma_s({r}_s)$ at site ${r}_s$ in the sensory layer, and spin $\sigma_h({r}_h)$ at site ${r}_h$ in the hidden layer, the approximate $\Delta E$ for a sensory spin $\sigma_s(r_s)$ experiencing an external applied field $h_e$ is: (note that  $\sigma = \pm 1$)
\begin{align}
\nonumber\Delta E(r_s) = &2\beta_s \Bigg\{ \Bigg. - m_s h_e\sigma_s (r_s) \\
&+m_s \sigma_s(r_s)\sum_{{r^{'}}_s} D(r_s-{r^{'}}_s)m_s\sigma_s({r^{'}}_s) + m_s \nonumber\sigma_s(r_s)\sum_{{r^{'}}_h} D(r_s-{r^{'}}_h)m_h\sigma_h({r^{'}}_h) \Bigg.\Bigg\} \\
	\label{eq:delta_s}\approx &2\beta_s\Bigg\{ \Bigg. -m_s h_e  \sigma_s (r_s)
+ m_s \sigma_s(r_s) \sum_{{r^{'}}_h}  D(r_s-{r^{'}}_h)m_h\sigma_h({r^{'}}_h)   \Bigg.\Bigg\}.
\end{align}

\noindent The first term in Eqn.~\ref{eq:delta_s} is the interaction energy of the sensory spins with the environment field and is proportional to $m_s$ of a sensory element. The second term represents the dipole field interactions between the sensory spins. 

The first dipole term in the top part of Eqn.~\eqref{eq:delta_s} is neglected under the assumption that the environment field energy is much larger than the dipole interaction between sensory spins. An additional assumption is that the sensory spin $m_s$ is much larger than the hidden spin $m_h$. This means the $h_e$ term dominates (except for very small applied fields) and the $D$ energy contributes effectively as a small noise term; it is neglected in the numerical calculations.

A spin in the hidden layer is described by the energy change:
\begin{align}
\nonumber\Delta E(r_h) = &2 \beta_h \Bigg\{ \Bigg. -m_h h_e  \sigma_h (r_h) \\
& +m_h\sigma_h (r_h)\sum_{{r^{'}}_h}  D_{\beta_h}(r_h-{r^{'}}_h)m_h\sigma_h({r^{'}}_h) + \nonumber m_h\sigma_h(r_h)\sum_{{r^{`}}_s}  D(r_h-{r^{'}}_s)m_s\sigma_s ({r^{'}}_s)    \Bigg.\Bigg\}\\
\label{eq:delta_h}\approx & 2\beta_h \Bigg\{ \Bigg. m_h\sigma_h(r_h) \sum_{{r^{'}}_h} D(r_h-{r^{'}}_h)m_h\sigma_h ({r^{'}}_h)+m_h\sigma_h(r_h)\sum_{r_s}  D(r_s-{r^{'}}_h)m_s\sigma_s ({r^{'}}_s)   \Bigg.\Bigg\}
\end{align}

\noindent Here, the first term in Eqn.~\ref{eq:delta_h} is the effect of the external environment field $h_e$ on the hidden spins. The second term represents interactions between all spins in the hidden layer. This external field is assumed to be applied locally to the sensor spins and is much weaker than the hidden spin interaction terms and therefore neglected. The third term represents the fields generated by sensory spins acting on the hidden layer spins.

An important aspect of this model is that because the effective temperatures for the sensory and hidden layer spins are different, timescales for reversal dynamics are also different. These are modelled such that the sensory spins relax with respect to the instantaneous value of the external field, while the hidden spins relax in accord to the fields generated by the sensory spins in their states. The condition $T_h > T_s$ describes sensory spins that sample the local input field less frequently than the hidden spins sample the sensory spin configuration. We find in the active inference experiments discussed above that the ratio of these temperatures affects the sampling of ASI configuration states.


\begin{acknowledgement}

The authors thank  M. Falconbridge for helpful and insightful discussions. This work was supported from the University of Manitoba, the Natural Sciences and Engineering Research Council of Canada (NSERC RGPIN 05011-18 and RGPIN-2018-05012), and the Canadian Foundation for Innovation (CFI) John R. Evans Leaders Fund.


\end{acknowledgement}

%
%
%

\bibliography{ASI_References_jvl.bib}

\providecommand{\latin}[1]{#1}
\makeatletter
\providecommand{\doi}
  {\begingroup\let\do\@makeother\dospecials
  \catcode`\{=1 \catcode`\}=2 \doi@aux}
\providecommand{\doi@aux}[1]{\endgroup\texttt{#1}}
\makeatother
\providecommand*\mcitethebibliography{\thebibliography}
\csname @ifundefined\endcsname{endmcitethebibliography}
  {\let\endmcitethebibliography\endthebibliography}{}
\begin{mcitethebibliography}{48}
\providecommand*\natexlab[1]{#1}
\providecommand*\mciteSetBstSublistMode[1]{}
\providecommand*\mciteSetBstMaxWidthForm[2]{}
\providecommand*\mciteBstWouldAddEndPuncttrue
  {\def\EndOfBibitem{\unskip.}}
\providecommand*\mciteBstWouldAddEndPunctfalse
  {\let\EndOfBibitem\relax}
\providecommand*\mciteSetBstMidEndSepPunct[3]{}
\providecommand*\mciteSetBstSublistLabelBeginEnd[3]{}
\providecommand*\EndOfBibitem{}
\mciteSetBstSublistMode{f}
\mciteSetBstMaxWidthForm{subitem}{(\alph{mcitesubitemcount})}
\mciteSetBstSublistLabelBeginEnd
  {\mcitemaxwidthsubitemform\space}
  {\relax}
  {\relax}

\bibitem[Friston \latin{et~al.}(2006)Friston, Kilner, and
  Harrison]{friston_free_2006}
Friston,~K.; Kilner,~J.; Harrison,~L. A free energy principle for the brain.
  \emph{Journal of Physiology-Paris} \textbf{2006}, \emph{100}, 70--87\relax
\mciteBstWouldAddEndPuncttrue
\mciteSetBstMidEndSepPunct{\mcitedefaultmidpunct}
{\mcitedefaultendpunct}{\mcitedefaultseppunct}\relax
\EndOfBibitem
\bibitem[Adams \latin{et~al.}(2015)Adams, Aponte, Marshall, and
  Friston]{adams_active_2015}
Adams,~R.~A.; Aponte,~E.; Marshall,~L.; Friston,~K.~J. Active inference and
  oculomotor pursuit: {The} dynamic causal modelling of eye movements.
  \emph{Journal of Neuroscience Methods} \textbf{2015}, \emph{242}, 1--14\relax
\mciteBstWouldAddEndPuncttrue
\mciteSetBstMidEndSepPunct{\mcitedefaultmidpunct}
{\mcitedefaultendpunct}{\mcitedefaultseppunct}\relax
\EndOfBibitem
\bibitem[Friston \latin{et~al.}(2008)Friston, Trujillo-Barreto, and
  Daunizeau]{friston_dem_2008}
Friston,~K.; Trujillo-Barreto,~N.; Daunizeau,~J. {DEM}: {A} variational
  treatment of dynamic systems. \emph{NeuroImage} \textbf{2008}, \emph{41},
  849--885\relax
\mciteBstWouldAddEndPuncttrue
\mciteSetBstMidEndSepPunct{\mcitedefaultmidpunct}
{\mcitedefaultendpunct}{\mcitedefaultseppunct}\relax
\EndOfBibitem
\bibitem[Friston \latin{et~al.}(2010)Friston, Stephan, Li, and
  Daunizeau]{friston_generalised_2010}
Friston,~K.; Stephan,~K.; Li,~B.; Daunizeau,~J. Generalised {Filtering}.
  \emph{Mathematical Problems in Engineering} \textbf{2010}, \emph{2010},
  1--34\relax
\mciteBstWouldAddEndPuncttrue
\mciteSetBstMidEndSepPunct{\mcitedefaultmidpunct}
{\mcitedefaultendpunct}{\mcitedefaultseppunct}\relax
\EndOfBibitem
\bibitem[Friston(2010)]{friston_free-energy_2010}
Friston,~K. The free-energy principle: a unified brain theory? \emph{Nat Rev
  Neurosci} \textbf{2010}, \emph{11}, 127--138\relax
\mciteBstWouldAddEndPuncttrue
\mciteSetBstMidEndSepPunct{\mcitedefaultmidpunct}
{\mcitedefaultendpunct}{\mcitedefaultseppunct}\relax
\EndOfBibitem
\bibitem[Friston(2009)]{friston_free-energy_2009}
Friston,~K. The free-energy principle: a rough guide to the brain? \emph{Trends
  in Cognitive Sciences} \textbf{2009}, \emph{13}, 293--301\relax
\mciteBstWouldAddEndPuncttrue
\mciteSetBstMidEndSepPunct{\mcitedefaultmidpunct}
{\mcitedefaultendpunct}{\mcitedefaultseppunct}\relax
\EndOfBibitem
\bibitem[Aguilera \latin{et~al.}(2022)Aguilera, Millidge, Tschantz, and
  Buckley]{aguilera_how_2022}
Aguilera,~M.; Millidge,~B.; Tschantz,~A.; Buckley,~C.~L. How particular is the
  physics of the free energy principle? \emph{Physics of Life Reviews}
  \textbf{2022}, \emph{40}, 24--50\relax
\mciteBstWouldAddEndPuncttrue
\mciteSetBstMidEndSepPunct{\mcitedefaultmidpunct}
{\mcitedefaultendpunct}{\mcitedefaultseppunct}\relax
\EndOfBibitem
\bibitem[FitzGerald \latin{et~al.}(2015)FitzGerald, Schwartenbeck, Moutoussis,
  Dolan, and Friston]{fitzgerald_active_2015}
FitzGerald,~T. H.~B.; Schwartenbeck,~P.; Moutoussis,~M.; Dolan,~R.~J.;
  Friston,~K. Active {Inference}, {Evidence} {Accumulation}, and the {Urn}
  {Task}. \emph{Neural Computation} \textbf{2015}, \emph{27}, 306--328\relax
\mciteBstWouldAddEndPuncttrue
\mciteSetBstMidEndSepPunct{\mcitedefaultmidpunct}
{\mcitedefaultendpunct}{\mcitedefaultseppunct}\relax
\EndOfBibitem
\bibitem[Nakajima(2020)]{Nakajima.2020}
Nakajima,~K. Physical reservoir computing—an introductory perspective.
  \emph{Japanese Journal of Applied Physics} \textbf{2020}, \emph{59},
  060501\relax
\mciteBstWouldAddEndPuncttrue
\mciteSetBstMidEndSepPunct{\mcitedefaultmidpunct}
{\mcitedefaultendpunct}{\mcitedefaultseppunct}\relax
\EndOfBibitem
\bibitem[Skjærvø \latin{et~al.}(2020)Skjærvø, Marrows, Stamps, and
  Heyderman]{skjaervo_advances_2020}
Skjærvø,~S.~H.; Marrows,~C.~H.; Stamps,~R.~L.; Heyderman,~L.~J. Advances in
  artificial spin ice. \emph{Nature Reviews Physics} \textbf{2020}, \emph{2},
  13--28\relax
\mciteBstWouldAddEndPuncttrue
\mciteSetBstMidEndSepPunct{\mcitedefaultmidpunct}
{\mcitedefaultendpunct}{\mcitedefaultseppunct}\relax
\EndOfBibitem
\bibitem[Nisoli \latin{et~al.}(2013)Nisoli, Moessner, and
  Schiffer]{nisoli_colloquium_2013}
Nisoli,~C.; Moessner,~R.; Schiffer,~P. Colloquium: {Artificial} spin ice:
  {Designing} and imaging magnetic frustration. \emph{Reviews of Modern
  Physics} \textbf{2013}, \emph{85}, 1473--1490\relax
\mciteBstWouldAddEndPuncttrue
\mciteSetBstMidEndSepPunct{\mcitedefaultmidpunct}
{\mcitedefaultendpunct}{\mcitedefaultseppunct}\relax
\EndOfBibitem
\bibitem[Marrows(2021)]{marrows_experimental_2021}
Marrows,~C.~H. In \emph{Spin {Ice}}; Udagawa,~M., Jaubert,~L., Eds.; Springer
  {Series} in {Solid}-{State} {Sciences}; Springer International Publishing:
  Cham, 2021; pp 455--478\relax
\mciteBstWouldAddEndPuncttrue
\mciteSetBstMidEndSepPunct{\mcitedefaultmidpunct}
{\mcitedefaultendpunct}{\mcitedefaultseppunct}\relax
\EndOfBibitem
\bibitem[Jensen and Tufte(2020)Jensen, and Tufte]{jensen_reservoir_2020}
Jensen,~J.~H.; Tufte,~G. Reservoir {Computing} in {Artificial} {Spin} {Ice}.
  2020; pp 376--383\relax
\mciteBstWouldAddEndPuncttrue
\mciteSetBstMidEndSepPunct{\mcitedefaultmidpunct}
{\mcitedefaultendpunct}{\mcitedefaultseppunct}\relax
\EndOfBibitem
\bibitem[Hon \latin{et~al.}(2021)Hon, Kuwabiraki, Goto, Nakatani, Suzuki, and
  Nomura]{hon_numerical_2021}
Hon,~K.; Kuwabiraki,~Y.; Goto,~M.; Nakatani,~R.; Suzuki,~Y.; Nomura,~H.
  Numerical simulation of artificial spin ice for reservoir computing.
  \emph{Applied Physics Express} \textbf{2021}, \emph{14}, 033001\relax
\mciteBstWouldAddEndPuncttrue
\mciteSetBstMidEndSepPunct{\mcitedefaultmidpunct}
{\mcitedefaultendpunct}{\mcitedefaultseppunct}\relax
\EndOfBibitem
\bibitem[Gartside \latin{et~al.}(2022)Gartside, Stenning, Vanstone, Holder,
  Arroo, Dion, Caravelli, Kurebayashi, and
  Branford]{gartside_reconfigurable_2022}
Gartside,~J.~C.; Stenning,~K.~D.; Vanstone,~A.; Holder,~H.~H.; Arroo,~D.~M.;
  Dion,~T.; Caravelli,~F.; Kurebayashi,~H.; Branford,~W.~R. Reconfigurable
  training and reservoir computing in an artificial spin-vortex ice via
  spin-wave fingerprinting. \emph{Nature Nanotechnology} \textbf{2022},
  \emph{17}, 460--469\relax
\mciteBstWouldAddEndPuncttrue
\mciteSetBstMidEndSepPunct{\mcitedefaultmidpunct}
{\mcitedefaultendpunct}{\mcitedefaultseppunct}\relax
\EndOfBibitem
\bibitem[May \latin{et~al.}(2021)May, Saccone, van~den Berg, Askey, Hunt, and
  Ladak]{may_magnetic_2021}
May,~A.; Saccone,~M.; van~den Berg,~A.; Askey,~J.; Hunt,~M.; Ladak,~S. Magnetic
  charge propagation upon a {3D} artificial spin-ice. \emph{Nature
  Communications} \textbf{2021}, \emph{12}, 3217\relax
\mciteBstWouldAddEndPuncttrue
\mciteSetBstMidEndSepPunct{\mcitedefaultmidpunct}
{\mcitedefaultendpunct}{\mcitedefaultseppunct}\relax
\EndOfBibitem
\bibitem[Saccone \latin{et~al.}(2023)Saccone, Van~den Berg, Harding, Singh,
  Giblin, Flicker, and Ladak]{saccone_exploring_2023}
Saccone,~M.; Van~den Berg,~A.; Harding,~E.; Singh,~S.; Giblin,~S.~R.;
  Flicker,~F.; Ladak,~S. Exploring the phase diagram of {3D} artificial
  spin-ice. \emph{Communications Physics} \textbf{2023}, \emph{6}, 1--9\relax
\mciteBstWouldAddEndPuncttrue
\mciteSetBstMidEndSepPunct{\mcitedefaultmidpunct}
{\mcitedefaultendpunct}{\mcitedefaultseppunct}\relax
\EndOfBibitem
\bibitem[Hopfield(1982)]{hopfield_neural_1982}
Hopfield,~J.~J. Neural networks and physical systems with emergent collective
  computational abilities. \emph{Proceedings of the National Academy of
  Sciences} \textbf{1982}, \emph{79}, 2554--2558, Publisher: Proceedings of the
  National Academy of Sciences\relax
\mciteBstWouldAddEndPuncttrue
\mciteSetBstMidEndSepPunct{\mcitedefaultmidpunct}
{\mcitedefaultendpunct}{\mcitedefaultseppunct}\relax
\EndOfBibitem
\bibitem[Coolen \latin{et~al.}(2005)Coolen, Kuehn, and
  Sollich]{coolen_theory_2005}
Coolen,~A. C.~C.; Kuehn,~R.; Sollich,~P. \emph{Theory of {Neural} {Information}
  {Processing} {Systems}}; Oxford University Press, Oxford, 2005\relax
\mciteBstWouldAddEndPuncttrue
\mciteSetBstMidEndSepPunct{\mcitedefaultmidpunct}
{\mcitedefaultendpunct}{\mcitedefaultseppunct}\relax
\EndOfBibitem
\bibitem[Begum~Popy \latin{et~al.}(2022)Begum~Popy, Frank, and
  Stamps]{begum_popy_magnetic_2022}
Begum~Popy,~R.; Frank,~J.; Stamps,~R.~L. Magnetic field driven dynamics in
  twisted bilayer artificial spin ice at superlattice angles. \emph{Journal of
  Applied Physics} \textbf{2022}, \emph{132}, 133902\relax
\mciteBstWouldAddEndPuncttrue
\mciteSetBstMidEndSepPunct{\mcitedefaultmidpunct}
{\mcitedefaultendpunct}{\mcitedefaultseppunct}\relax
\EndOfBibitem
\bibitem[Castelnovo \latin{et~al.}(2008)Castelnovo, Moessner, and
  Sondhi]{castelnovo_magnetic_2008}
Castelnovo,~C.; Moessner,~R.; Sondhi,~S.~L. Magnetic monopoles in spin ice.
  \emph{Nature} \textbf{2008}, \emph{451}, 42--45\relax
\mciteBstWouldAddEndPuncttrue
\mciteSetBstMidEndSepPunct{\mcitedefaultmidpunct}
{\mcitedefaultendpunct}{\mcitedefaultseppunct}\relax
\EndOfBibitem
\bibitem[Friston(2013)]{friston_life_2013}
Friston,~K. Life as we know it. \emph{Journal of The Royal Society Interface}
  \textbf{2013}, \emph{10}, 20130475\relax
\mciteBstWouldAddEndPuncttrue
\mciteSetBstMidEndSepPunct{\mcitedefaultmidpunct}
{\mcitedefaultendpunct}{\mcitedefaultseppunct}\relax
\EndOfBibitem
\bibitem[Kirchhoff \latin{et~al.}()Kirchhoff, Parr, Palacios, Friston, and
  Kiverstein]{kirchhoff_markov_2018}
Kirchhoff,~M.; Parr,~T.; Palacios,~E.; Friston,~K.; Kiverstein,~J. The Markov
  blankets of life: autonomy, active inference and the free energy principle.
  \emph{J. R. Soc. Interface.} \emph{15}, 20170792\relax
\mciteBstWouldAddEndPuncttrue
\mciteSetBstMidEndSepPunct{\mcitedefaultmidpunct}
{\mcitedefaultendpunct}{\mcitedefaultseppunct}\relax
\EndOfBibitem
\bibitem[Budrikis \latin{et~al.}(2011)Budrikis, Politi, and
  Stamps]{budrikis_diversity_2011}
Budrikis,~Z.; Politi,~P.; Stamps,~R.~L. Diversity {Enabling} {Equilibration}:
  {Disorder} and the {Ground} {State} in {Artificial} {Spin} {Ice}.
  \emph{Physical Review Letters} \textbf{2011}, \emph{107}, 217204, Publisher:
  American Physical Society\relax
\mciteBstWouldAddEndPuncttrue
\mciteSetBstMidEndSepPunct{\mcitedefaultmidpunct}
{\mcitedefaultendpunct}{\mcitedefaultseppunct}\relax
\EndOfBibitem
\bibitem[Budrikis \latin{et~al.}(2012)Budrikis, Morgan, Akerman, Stein, Politi,
  Langridge, Marrows, and Stamps]{budrikis_disorder_2012}
Budrikis,~Z.; Morgan,~J.~P.; Akerman,~J.; Stein,~A.; Politi,~P.; Langridge,~S.;
  Marrows,~C.~H.; Stamps,~R.~L. Disorder {Strength} and {Field}-{Driven}
  {Ground} {State} {Domain} {Formation} in {Artificial} {Spin} {Ice}:
  {Experiment}, {Simulation}, and {Theory}. \emph{Physical Review Letters}
  \textbf{2012}, \emph{109}, 037203\relax
\mciteBstWouldAddEndPuncttrue
\mciteSetBstMidEndSepPunct{\mcitedefaultmidpunct}
{\mcitedefaultendpunct}{\mcitedefaultseppunct}\relax
\EndOfBibitem
\bibitem[Friston and Stephan()Friston, and Stephan]{friston_free-energy_2007}
Friston,~K.~J.; Stephan,~K.~E. Free-energy and the brain. \emph{Synthese}
  \emph{159}, 417--458\relax
\mciteBstWouldAddEndPuncttrue
\mciteSetBstMidEndSepPunct{\mcitedefaultmidpunct}
{\mcitedefaultendpunct}{\mcitedefaultseppunct}\relax
\EndOfBibitem
\bibitem[Buckley \latin{et~al.}()Buckley, Kim, {McGregor}, and
  Seth]{buckley_free_2017}
Buckley,~C.~L.; Kim,~C.~S.; {McGregor},~S.; Seth,~A.~K. The free energy
  principle for action and perception: A mathematical review. \emph{Journal of
  Mathematical Psychology} \emph{81}, 55--79\relax
\mciteBstWouldAddEndPuncttrue
\mciteSetBstMidEndSepPunct{\mcitedefaultmidpunct}
{\mcitedefaultendpunct}{\mcitedefaultseppunct}\relax
\EndOfBibitem
\bibitem[Balaji()]{balaji_continuous-discrete_2009}
Balaji,~B. Continuous-Discrete Path Integral Filtering. \emph{Entropy}
  \emph{11}, 402--430\relax
\mciteBstWouldAddEndPuncttrue
\mciteSetBstMidEndSepPunct{\mcitedefaultmidpunct}
{\mcitedefaultendpunct}{\mcitedefaultseppunct}\relax
\EndOfBibitem
\bibitem[Baltieri(2019)]{baltieri_active_2019}
Baltieri,~M. \emph{Active {Inference}: {Building} a {New} {Bridge} {Between}
  {Control} {Theory} and {Embodied} {Cognitive} {Science}}; University of
  Sussex, 2019\relax
\mciteBstWouldAddEndPuncttrue
\mciteSetBstMidEndSepPunct{\mcitedefaultmidpunct}
{\mcitedefaultendpunct}{\mcitedefaultseppunct}\relax
\EndOfBibitem
\bibitem[Balaji and Friston()Balaji, and Friston]{balaji_bayesian_2011}
Balaji,~B.; Friston,~K. Bayesian state estimation using generalized
  coordinates. p 80501Y\relax
\mciteBstWouldAddEndPuncttrue
\mciteSetBstMidEndSepPunct{\mcitedefaultmidpunct}
{\mcitedefaultendpunct}{\mcitedefaultseppunct}\relax
\EndOfBibitem
\bibitem[Friston \latin{et~al.}()Friston, Mattout, Trujillo-Barreto, Ashburner,
  and Penny]{friston_variational_2007}
Friston,~K.; Mattout,~J.; Trujillo-Barreto,~N.; Ashburner,~J.; Penny,~W.
  Variational free energy and the Laplace approximation. \emph{{NeuroImage}}
  \emph{34}, 220--234\relax
\mciteBstWouldAddEndPuncttrue
\mciteSetBstMidEndSepPunct{\mcitedefaultmidpunct}
{\mcitedefaultendpunct}{\mcitedefaultseppunct}\relax
\EndOfBibitem
\bibitem[Baltieri(2020)]{baltieri_bayesian_2020}
Baltieri,~M. A {Bayesian} perspective on classical control. 2020
  {International} {Joint} {Conference} on {Neural} {Networks} ({IJCNN}). 2020;
  pp 1--8, ISSN: 2161-4407\relax
\mciteBstWouldAddEndPuncttrue
\mciteSetBstMidEndSepPunct{\mcitedefaultmidpunct}
{\mcitedefaultendpunct}{\mcitedefaultseppunct}\relax
\EndOfBibitem
\bibitem[Adams \latin{et~al.}(2016)Adams, Bauer, Pinotsis, and
  Friston]{adams_dynamic_2016}
Adams,~R.~A.; Bauer,~M.; Pinotsis,~D.; Friston,~K.~J. Dynamic causal modelling
  of eye movements during pursuit: {Confirming} precision-encoding in {V1}
  using {MEG}. \emph{NeuroImage} \textbf{2016}, \emph{132}, 175--189\relax
\mciteBstWouldAddEndPuncttrue
\mciteSetBstMidEndSepPunct{\mcitedefaultmidpunct}
{\mcitedefaultendpunct}{\mcitedefaultseppunct}\relax
\EndOfBibitem
\bibitem[Isomura \latin{et~al.}()Isomura, Kotani, Jimbo, and
  Friston]{isomura_experimental_2023}
Isomura,~T.; Kotani,~K.; Jimbo,~Y.; Friston,~K.~J. Experimental validation of
  the free-energy principle with in vitro neural networks. \emph{Nat Commun}
  \emph{14}, 4547\relax
\mciteBstWouldAddEndPuncttrue
\mciteSetBstMidEndSepPunct{\mcitedefaultmidpunct}
{\mcitedefaultendpunct}{\mcitedefaultseppunct}\relax
\EndOfBibitem
\bibitem[Baños \latin{et~al.}(2010)Baños, Cruz, Fernandez, Gil-Narvion,
  Gordillo-Guerrero, Guidetti, Maiorano, Mantovani, Marinari, Martin-Mayor,
  Monforte-Garcia, Sudupe, Navarro, Parisi, Perez-Gaviro, Ruiz-Lorenzo,
  Schifano, Seoane, Tarancon, Tripiccione, and Yllanes]{banos_nature_2010}
Baños,~R.~A.; Cruz,~A.; Fernandez,~L.~A.; Gil-Narvion,~J.~M.;
  Gordillo-Guerrero,~A.; Guidetti,~M.; Maiorano,~A.; Mantovani,~F.;
  Marinari,~E.; Martin-Mayor,~V.; Monforte-Garcia,~J.; Sudupe,~A.~M.;
  Navarro,~D.; Parisi,~G.; Perez-Gaviro,~S.; Ruiz-Lorenzo,~J.~J.;
  Schifano,~S.~F.; Seoane,~B.; Tarancon,~A.; Tripiccione,~R. \latin{et~al.}
  Nature of the spin-glass phase at experimental length scales. \emph{J. Stat.
  Mech.} \textbf{2010}, \emph{2010}, P06026\relax
\mciteBstWouldAddEndPuncttrue
\mciteSetBstMidEndSepPunct{\mcitedefaultmidpunct}
{\mcitedefaultendpunct}{\mcitedefaultseppunct}\relax
\EndOfBibitem
\bibitem[Koudahl and de~Vries()Koudahl, and de~Vries]{koudahl_worked_2020}
Koudahl,~M.~T.; de~Vries,~B. A Worked Example of Fokker-Planck-Based Active
  Inference. Active Inference. pp 28--34\relax
\mciteBstWouldAddEndPuncttrue
\mciteSetBstMidEndSepPunct{\mcitedefaultmidpunct}
{\mcitedefaultendpunct}{\mcitedefaultseppunct}\relax
\EndOfBibitem
\bibitem[Friston \latin{et~al.}(2023)Friston, Da~Costa, Sajid, Heins,
  Ueltzhöffer, Pavliotis, and Parr]{friston_free_2023}
Friston,~K.; Da~Costa,~L.; Sajid,~N.; Heins,~C.; Ueltzhöffer,~K.;
  Pavliotis,~G.~A.; Parr,~T. The free energy principle made simpler but not too
  simple. \emph{Physics Reports} \textbf{2023}, \emph{1024}, 1--29\relax
\mciteBstWouldAddEndPuncttrue
\mciteSetBstMidEndSepPunct{\mcitedefaultmidpunct}
{\mcitedefaultendpunct}{\mcitedefaultseppunct}\relax
\EndOfBibitem
\bibitem[Heins \latin{et~al.}(2023)Heins, Klein, Demekas, Aguilera, and
  Buckley]{heins_spin_2023}
Heins,~C.; Klein,~B.; Demekas,~D.; Aguilera,~M.; Buckley,~C.~L. Spin {Glass}
  {Systems} as {Collective} {Active} {Inference}. Active {Inference}. Cham,
  2023; pp 75--98\relax
\mciteBstWouldAddEndPuncttrue
\mciteSetBstMidEndSepPunct{\mcitedefaultmidpunct}
{\mcitedefaultendpunct}{\mcitedefaultseppunct}\relax
\EndOfBibitem
\bibitem[Nickel \latin{et~al.}(2023)Nickel, Gibbs, Burgess, Shafer, {Motta
  Meira}, Sun, and van Lierop]{Nickel.2023}
Nickel,~R.; Gibbs,~J.; Burgess,~J.; Shafer,~P.; {Motta Meira},~D.; Sun,~C.; van
  Lierop,~J. Nanoscale size effects on push-pull Fe-O hybridization through the
  multiferroic transition of perovskite {$\epsilon$-Fe$_2$O$_3$}. \emph{Nano
  Letters} \textbf{2023}, \emph{23}, 7845\relax
\mciteBstWouldAddEndPuncttrue
\mciteSetBstMidEndSepPunct{\mcitedefaultmidpunct}
{\mcitedefaultendpunct}{\mcitedefaultseppunct}\relax
\EndOfBibitem
\bibitem[Torrejon \latin{et~al.}(2017)Torrejon, Riou, Araujo, Tsunegi, Khalsa,
  Querlioz, Bortolotti, Cros, Yakushiji, Fukushima, Kubota, Yuasa, Stiles, and
  Grollier]{torrejon_neuromorphic_2017-1}
Torrejon,~J.; Riou,~M.; Araujo,~F.~A.; Tsunegi,~S.; Khalsa,~G.; Querlioz,~D.;
  Bortolotti,~P.; Cros,~V.; Yakushiji,~K.; Fukushima,~A.; Kubota,~H.;
  Yuasa,~S.; Stiles,~M.~D.; Grollier,~J. Neuromorphic computing with nanoscale
  spintronic oscillators. \emph{Nature} \textbf{2017}, \emph{547},
  428--431\relax
\mciteBstWouldAddEndPuncttrue
\mciteSetBstMidEndSepPunct{\mcitedefaultmidpunct}
{\mcitedefaultendpunct}{\mcitedefaultseppunct}\relax
\EndOfBibitem
\bibitem[Grollier \latin{et~al.}(2020)Grollier, Querlioz, Camsari,
  Everschor-Sitte, Fukami, and Stiles]{grollier_neuromorphic_2020}
Grollier,~J.; Querlioz,~D.; Camsari,~K.~Y.; Everschor-Sitte,~K.; Fukami,~S.;
  Stiles,~M.~D. Neuromorphic spintronics. \emph{Nature Electronics}
  \textbf{2020}, \emph{3}, 360--370\relax
\mciteBstWouldAddEndPuncttrue
\mciteSetBstMidEndSepPunct{\mcitedefaultmidpunct}
{\mcitedefaultendpunct}{\mcitedefaultseppunct}\relax
\EndOfBibitem
\bibitem[Lee and Mumford(2003)Lee, and Mumford]{lee_hierarchical_2003}
Lee,~T.~S.; Mumford,~D. Hierarchical {Bayesian} inference in the visual cortex.
  \emph{J. Opt. Soc. Am. A} \textbf{2003}, \emph{20}, 1434\relax
\mciteBstWouldAddEndPuncttrue
\mciteSetBstMidEndSepPunct{\mcitedefaultmidpunct}
{\mcitedefaultendpunct}{\mcitedefaultseppunct}\relax
\EndOfBibitem
\bibitem[Budrikis(2014)]{BUDRIKIS2014109}
Budrikis,~Z. In \emph{Chapter Two - Disorder, Edge, and Field Protocol Effects
  in Athermal Dynamics of Artificial Spin Ice}; Camley,~R.~E., Stamps,~R.~L.,
  Eds.; Solid State Physics; Academic Press, 2014; Vol.~65; pp 109--236\relax
\mciteBstWouldAddEndPuncttrue
\mciteSetBstMidEndSepPunct{\mcitedefaultmidpunct}
{\mcitedefaultendpunct}{\mcitedefaultseppunct}\relax
\EndOfBibitem
\bibitem[Desplat \latin{et~al.}(2018)Desplat, Suess, Kim, and
  Stamps]{desplat_thermal_2018}
Desplat,~L.; Suess,~D.; Kim,~J.-V.; Stamps,~R.~L. Thermal stability of
  metastable magnetic skyrmions: {Entropic} narrowing and significance of
  internal eigenmodes. \emph{Physical Review B} \textbf{2018}, \emph{98},
  134407\relax
\mciteBstWouldAddEndPuncttrue
\mciteSetBstMidEndSepPunct{\mcitedefaultmidpunct}
{\mcitedefaultendpunct}{\mcitedefaultseppunct}\relax
\EndOfBibitem
\bibitem[Desplat \latin{et~al.}(2019)Desplat, Kim, and
  Stamps]{desplat_paths_2019}
Desplat,~L.; Kim,~J.-V.; Stamps,~R.~L. Paths to annihilation of first- and
  second-order (anti)skyrmions via (anti)meron nucleation on the frustrated
  square lattice. \emph{Physical Review B} \textbf{2019}, \emph{99},
  174409\relax
\mciteBstWouldAddEndPuncttrue
\mciteSetBstMidEndSepPunct{\mcitedefaultmidpunct}
{\mcitedefaultendpunct}{\mcitedefaultseppunct}\relax
\EndOfBibitem
\bibitem[Pinna \latin{et~al.}(2020)Pinna, Bourianoff, and
  Everschor-Sitte]{PhysRevApplied.14.054020}
Pinna,~D.; Bourianoff,~G.; Everschor-Sitte,~K. Reservoir Computing with Random
  Skyrmion Textures. \emph{Phys. Rev. Appl.} \textbf{2020}, \emph{14},
  054020\relax
\mciteBstWouldAddEndPuncttrue
\mciteSetBstMidEndSepPunct{\mcitedefaultmidpunct}
{\mcitedefaultendpunct}{\mcitedefaultseppunct}\relax
\EndOfBibitem
\bibitem[Raab \latin{et~al.}(2022)Raab, Brems, Beneke, Dohi, Rothörl,
  Kammerbauer, Mentink, and Kläui]{raab_brownian_2022}
Raab,~K.; Brems,~M.~A.; Beneke,~G.; Dohi,~T.; Rothörl,~J.; Kammerbauer,~F.;
  Mentink,~J.~H.; Kläui,~M. Brownian reservoir computing realized using
  geometrically confined skyrmion dynamics. \emph{Nat Commun} \textbf{2022},
  \emph{13}, 6982\relax
\mciteBstWouldAddEndPuncttrue
\mciteSetBstMidEndSepPunct{\mcitedefaultmidpunct}
{\mcitedefaultendpunct}{\mcitedefaultseppunct}\relax
\EndOfBibitem
\end{mcitethebibliography}

\end{document}